\documentclass[letter, 10pt, conference]{IEEEtran}
\usepackage{amsmath}
\usepackage{amssymb}
\usepackage{amsfonts}
\usepackage{graphicx}
\usepackage{float}
\usepackage{graphics}
\usepackage{graphicx}
\usepackage{theorem}

\IEEEoverridecommandlockouts
\begin{document}

\title{{\LARGE \textbf{Capacity of Linear Two-hop Mesh Networks with Rate Splitting,
Decode-and-forward Relaying and Cooperation}}}
\author{O. Simeone, O. Somekh, Y. Bar-Ness, H. V. Poor, and S. Shamai (Shitz)%
\thanks{%
O. Simeone and Y. Bar-Ness are with the Center for Wireless Communications
and Signal Processing Research, New Jersey Institute of Technology,\ Newark,
New Jersey 07102-1982, USA\texttt{\ {\small \{osvaldo.simeone@njit.edu,
barness@yegal.njit.edu\}.}}}\thanks{%
O. Somekh and H. V. Poor are with the Department of Electrical Engineering,
Princeton University, Princeton, NJ 08544, USA \texttt{\small %
\{orens@princeton.edu, poor@princeton.edu\}.}}\thanks{%
S. Shamai (Shitz) is with the Department of Electrical Engineering,
Technion, Haifa, 32000 \texttt{\small \{sshlomo@ee.technion.ac.il\}.} }%
\thanks{%
This research was supported in part by the U.S. National Science
Foundation under Grants CNS-06-26611 and CNS-06-25637, and by a
Marie Curie Outgoing International Fellowship within the 6th
European Community Framework Programme.}} \maketitle

\begin{abstract}
A linear mesh network is considered in which a single user per cell
communicates to a local base station via a dedicated relay (two-hop
communication). Exploiting the possibly relevant inter-cell channel gains,
rate splitting with successive cancellation in both hops is investigated as
a promising solution to improve the rate of basic single-rate
communications. Then, an alternative solution is proposed that attempts to
improve the performance of the second hop (from the relays to base stations)
by cooperative transmission among the relay stations. The cooperative scheme
leverages the common information obtained by the relays as a by-product of
the use of rate splitting in the first hop. Numerical results bring insight
into the conditions (network topology and power constraints) under which
rate splitting, with possible relay cooperation, is beneficial. Multi-cell
processing (joint decoding at the base stations) is also considered for
reference.
\end{abstract}

\thispagestyle{empty} \pagestyle{empty}


\section{Introduction}

\thispagestyle{empty}

Wireless mesh networks are currently being investigated for their potential
to resolve the performance limitations of both infrastructure (cellular) and
multi-hop (ad hoc) networks in terms of quality-of-service and coverage \cite%
{mesh}. Basically, mesh networks prescribe the combination of communication
via direct transmission to infrastructure nodes (base stations) and via
multi-hop transmission through intermediate nodes (relay stations). The
latter can generally be mobile terminals, or fixed stations appropriately
located by the service provider. The assessment of the performance of such
networks is an open problem that has attracted interest from different
communities and fields, especially information-theory \cite{infocom} \cite%
{hybrid} and networking \cite{networking}. Recently, there has
also been considerable interest in further enhancing the
performance of infrastructure or mesh networks by endowing the
system with a central processor able to pool the signals received
by the base stations and perform joint processing (this scenario
is usually referred to as distributed antennas or multi-cell
processing) \cite{somekh-review}.

In this paper, we focus on a linear mesh network as sketched in Fig. \ref%
{topology1}. It is assumed that one\ mobile terminal (MT)\ is active in each
cell in a given time-frequency resource (as for intra-cell TDMA\ or FDMA)
and that each active MT communicates with the same-cell base station (BS)
via a dedicated relay station (RS) (two-hop transmission). In order to allow
meaningful analysis and insight, this scenario is modelled as illustrated in
Fig. \ref{topology2}, where symmetry is assumed in the channel gains, i.e.,
every cell is characterized by identical intra- and inter-cell propagation
conditions. This framework follows the approach of \cite{simeone twc} (see
also \cite{somekh-review}), which extends the model of \cite{wyner} to mesh
networks.

The basic premise of this work is that the model in Fig. \ref{topology2} can
be seen as the cascade of two interference channels, one for each hop, with
many sources and corresponding receivers (border effects are neglected).
Therefore, from the literature on interference channels, a promising
approach is that of employing rate splitting with successive interference
cancellation at the receivers \cite{han} \cite{kramer}. It is recalled that
the rationale of rate splitting is that joint decoding of (at least part of)
the transmitted signals at the receivers has the potential to improve the
achievable rates with respect to single-user decoders that treat signals
other than the intended as noise. The main contributions of this work
concerning the analysis of a mesh network modelled as in Fig. \ref{topology2}
are:

\begin{itemize}
\item derivation of the performance of rate splitting applied to both hops
with decode-and-forward relaying (Sec. \ref{sec_transmission});

\item proposal of a cooperative transmission scheme for the RSs that
leverages the common information obtained by the relays as a by-product of
the use of rate splitting in the first hop (Sec. \ref{sec_second});

\item analysis of the cooperative transmission scheme above in the presence
of multi-cell processing (Sec. \ref{sec_second}); and

\item performance evaluation of rate splitting, with possible relay
cooperation in the second hop, via numerical simulations; comparison with
the reference cases of single-rate transmission and multi-cell processing is
provided as well (Sec. \ref{sec_numerical}).
\end{itemize}

Related work was recently reported in \cite{simeone twc} \cite{somekh} \cite%
{simeone-df}, where a cellular model similar to the one in Fig. \ref%
{topology2} was addressed under the assumption of amplify-and-forward \cite%
{simeone twc} \cite{somekh} or decode-and-forward (DF)\ relaying \cite%
{simeone-df} with single-rate transmission.

\section{System model \label{introduction}}

\begin{figure}
\centering
 \includegraphics[scale=0.42]{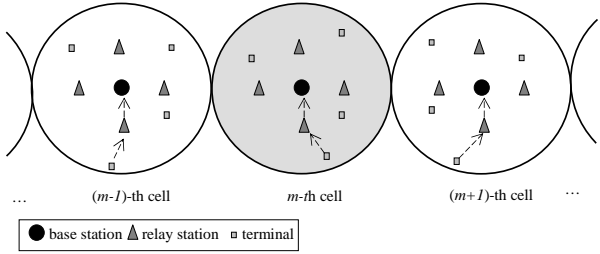}
 \caption{A linear two-hop mesh network.}
 \label{topology1}
\end{figure}

\begin{figure}
\centering
 \includegraphics[scale=0.42]{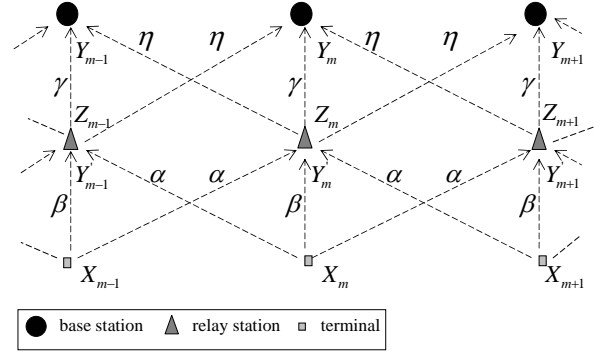}
 \caption{A schematic model of the linear two-hop mesh network.}
 \label{topology2}
\end{figure}

We study the abstraction of the two-hop mesh network of Fig. \ref{topology1}
as sketched in Fig. \ref{topology2}. Cells are arranged in a linear fashion,
one user transmitting on a given time-frequency resource in each cell.
Moreover, we focus on non-faded Gaussian channels and assume homogeneous
conditions for the channel power gains so that the intra-cell MS-to-RS
(first hop) and RS-to-BS (second hop) power gains are $\beta ^{2}$ and $%
\gamma ^{2},$ respectively, for all cells, and, similarly, the inter-cell
power gains between adjacent cells are $\alpha ^{2}\leq \beta ^{2}$ and $%
\eta ^{2}\leq \gamma ^{2}$ for first and second hop, respectively. Notice
that as in \cite{wyner} each cell receives signals only from adjacent cells.
Moreover, here there exist no direct paths between MTs and BSs and no
relevant inter-channels between RSs in adjacent cells. Because of the latter
assumptions, we can deal with either full duplex or half duplex transmission
at the relays with minor modifications, as explained below. Considering, for
simplicity of exposition, full-duplex transmission (by means of perfect
echo-cancellation), the signal received at each time by the $m$th RS (first
hop) can be written as
\begin{equation}
Y_{m}^{\prime }=\beta X_{m}+\alpha (X_{m-1}+X_{m+1})+N_{m},
\label{rx signal relay}
\end{equation}%
where $\beta $ and $\alpha $ are the (real) channel gains, and we assume the
symbols transmitted by the MTs, $X_{m},$ to be drawn from a circularly
symmetric complex Gaussian distribution with power $E[|X_{m}|^{2}]=P_{1}.$
Moreover, the additive noise $N_{m}$ is complex Gaussian with $%
E[|N_{m}|^{2}]=1.$ Similarly, the signal received by the $m$th BS is
\begin{equation}
Y_{m}=\gamma Z_{m}+\eta (Z_{m-1}+Z_{m+1})+M_{m},  \label{rx signal bs}
\end{equation}%
where the symbols transmitted by the RSs satisfy $E[|Z_{m}|^{2}]=P_{2}$ and
the additive Gaussian noise is such that $E[|M_{m}|^{2}]=1.$

By symmetry, we are interested in evaluating the common rate achievable by
all of the MTs over the network described by Fig. \ref{topology2} and
equations (\ref{rx signal relay})-(\ref{rx signal bs}). In order to simplify
the treatment, we will assume that the number of cells is large enough in
order to neglect border effects (see \cite{somekh-review} for further
discussion on this point in the context of the cellular model of \cite{wyner}%
).

\section{Achievable rate with rate splitting\label{sec_transmission}}

As mentioned above, in this paper we focus for simplicity of exposition on
full-duplex RSs. Accordingly, we assume a delayed block-by-block
transmission strategy whereby the information is transmitted through
multiple blocks, and the number of blocks is large enough so that we can
neglect the loss in spectral efficiency associated with the transmission of
first (MT\ to RS) and last (RS to BS) blocks. More specifically, in each
block, the MTs communicate new information to the RSs, and, at the same
time, the RSs forward (after decoding) the information received in the
previous block to the BSs. The absence of a direct path between MTs and BSs
allows RSs and BSs to perform block-by-block decoding without resorting to
more complicated decoding strategies \cite{kramer}. Moreover, for the same
reason, the full-duplex coding schemes considered in this paper can be
easily adapted to half-duplex RSs by simply alternating transmission from MT
or RS in each block. In the case of half-duplex then, since the MTs transmit
new information only once every two blocks, the corresponding achievable
rates are easily seen to be just half of the corresponding rates with full
duplex derived here\footnote{%
Strictly speaking, under average power constraint, the power used with
half-duplex by both sources and relays can be doubled with respect to the
full-duplex case.}.

In this section, we first review the basic reference case of single-rate
transmission (Sec. \ref{sec_single rate}) and then evaluate the achievable
rate with rate splitting in both hops (Sec. \ref{sec_rs_region} and \ref%
{sec_rs}).

\subsection{The reference case: single-rate transmission\label{sec_single
rate}}

As a preliminary example and reference case, consider the following simple
coding scheme based on DF relaying (further analyzed in a more general
framework in \cite{simeone-df}). In every block, each MT\ transmits to the
same-cell RS a Gaussian codeword taken from a rate-$R$ codebook. The RS
decodes the message treating the signals from adjacent cells as Gaussian
interference (single-user decoding), and forwards it in the next block to
the same-cell BS, that finally performs single-user decoding. The maximum
achievable rate per user of this scheme is easily shown to be%
\begin{equation}
R_{o}=\mathrm{C}\left( \min \left( \frac{\beta
^{2}P_{1}}{1+2\alpha ^{2}P_{1}},\frac{\gamma ^{2}P_{2}}{1+2\eta
^{2}P_{2}}\right) \right) , \label{Ro}
\end{equation}%
where we have defined the function $\mathrm{C}(x)=\log (1+x)$ and
the two terms inside the inner parentheses correspond to the
signal-to-interference-plus-noise ratios (SINRs) at the RS and BS,
respectively. The performance of this scheme is poor when the
inter-cell interference, i.e., the value of parameters $\alpha
^{2}$ and $\eta ^{2},$ is large. In the next section, we attempt
to alleviate this problem by leveraging on the idea of rate
splitting with Multiple Access Channel (MAC) decomposition, first
employed in \cite{han} in the context of the conventional
($2\times 2)$ interference channel (see also \cite{kramer}).

\subsection{Rate splitting for transmission to the RSs\label{sec_rs_region}}

In this section, we focus on the first hop, between MTs and RSs, and propose
a coding scheme based on the principle of rate splitting for the
interference channel \cite{han}. Accordingly, each MT transmits the sum of
two random Gaussian codebooks,
\begin{equation}
X_{m}=X_{p,m}(W_{p,m})+X_{c,m}(W_{c,m})\text{:}  \label{Xm_rs}
\end{equation}%
a \textit{private }codebook $X_{p,m}(\cdot )$ encoding a message $W_{p,m}$
intended to be decoded only at the same-cell RS, and a \textit{common}
codebook $X_{c,m}(\cdot )$ that carries a message $W_{c,m}$ to be decoded
not only at the same-cell RS but also at the two adjacent-cell RSs\footnote{%
Notice that the definition of private and common messages here is
receiver-centric, whereas elsewhere (see, e.g., \cite{slepian} \cite{maric1}
\cite{jiang}) it refers to the message availability at the transmitters (but
see also Sec. \ref{sec_coop}).}. The rate of the private and common
codebooks are denoted as $R_{1p}$ and $R_{1c},$ respectively (i.e., $%
W_{p,m}\in \{1,2,...,2^{nR_{1p}}\}$ and $W_{c,m}\in \{1,2,...,2^{nR_{1c}}\})$%
, whereas the corresponding powers are $P_{1p}=E[|X_{p,m}|^{2}]$ and $%
P_{1c}=[|X_{c,m}|^{2}]$. The total power per MT $P_{1}$ is divided among the
two codebooks as $P_{1}=P_{1p}+P_{1c}.$ Similarly, the total rate
transmitted by the user to the same-cell RS\ is given by $%
R_{rs,1}=R_{1p}+R_{1c}.$ Notice that each RS is informed of the private
codebook used by the same-cell MT and of the common codebooks employed by
the same-cell MTs and the two adjacent-cell MTs.

\begin{figure}
\centering
 \includegraphics[scale=0.6]{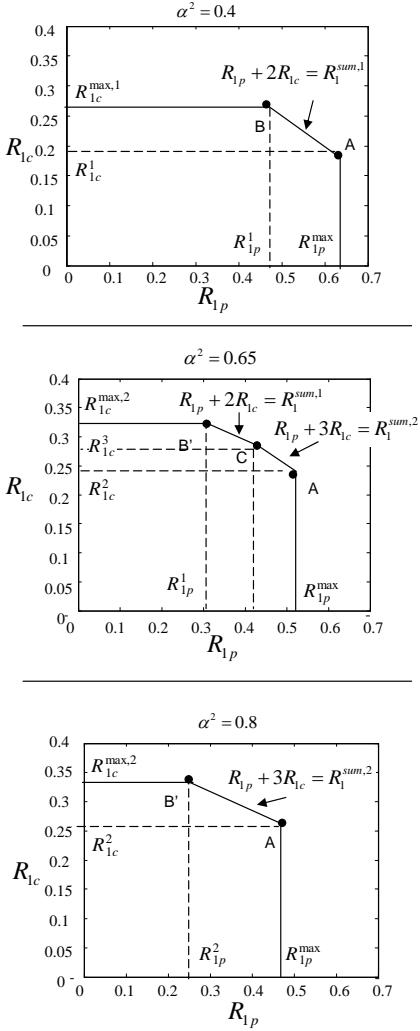}
 \caption{Three illustrative cases for the
capacity region (in terms of rates of private message, $R_{1p},$
and common message, $R_{1c})$ of rate splitting on the first hop,
corresponding to
different values of the inter-cell power gain $\protect\alpha ^{2}$ ($%
P_{1p}=1,$ $P_{1c}=1,$ $\protect\beta ^{2}=1).$ The
rate-maximizing vertex A is achievable by successive interference
cancellation where common messages are decoded first followed by
the same-cell private message. }
 \label{achievableregions}
\end{figure}

From (\ref{rx signal relay}) and (\ref{Xm_rs}), the signal received at each $%
m$th RS can be written as (dropping the arguments of the codewords):%
\begin{eqnarray}
Y_{m}^{\prime } &=&\beta (X_{p,m}+X_{c,m})+\alpha (X_{c,m-1}+X_{c,m+1})+
\label{rx signal rs} \\
&&+S_{m}+N_{m},  \notag
\end{eqnarray}%
where
\begin{equation}
S_{m}=\alpha (X_{p,m-1}+X_{p,m+1}).  \label{Sm1}
\end{equation}%
Based on (\ref{rx signal rs}), we assume that each $m$th RS jointly decodes
four messages: the private message $W_{p,m}$\ and the common message $%
W_{c,m} $ of the same-cell MT, and the common messages $W_{c,m-1}$ and $%
W_{c,m+1}$ of the two adjacent-cell MTs. The private messages $W_{p,m-1}$
and $W_{p,m+1} $ of the two adjacent-cell MTs are instead considered as the
(Gaussian) interference terms $S_{m}$ (\ref{Sm1}) with power $%
E[|S_{m}|^{2}]=2\alpha ^{2}P_{1p}.$ The channel (\ref{rx signal rs}) seen at
any $m$th RS\ is then a four-user MAC with inputs $X_{p,m},$ $X_{c,m},$ $%
X_{c,m-1}$ and $X_{c,m+1}$ and equivalent Gaussian noise with power $%
1+2\alpha ^{2}P_{1p}.$ Accordingly, for each choice of the power allocation (%
$P_{1p},P_{1c}$), the achievable rates $R_{1p}$ and $R_{1c}$ are limited by
the fifteen inequalities defining the capacity region $\mathcal{R}%
_{rs,1}(P_{1p},P_{1c})$ of the Gaussian MAC\ at hand \cite{cover}, which are
easily shown to boil down to:
\begin{subequations}
\label{Rrs1_region}
\begin{eqnarray}
R_{1p} &\leq &\mathrm{C}\left( \frac{\beta ^{2}P_{1p}}{1+2\alpha ^{2}P_{1p}}%
\right) \triangleq R_{1p}^{\max }(P_{1p})  \label{R1pmax} \\
R_{1c} &\leq &\min \left\{ \frac{1}{2}\mathrm{C}\left(
\frac{2\alpha
^{2}P_{1c}}{1+2\alpha ^{2}P_{1p}}\right) ,\right. \\
&&\left. \frac{1}{3}\mathrm{C}\left( \frac{(2\alpha ^{2}+\beta ^{2})P_{1c}}{%
1+2\alpha ^{2}P_{1p}}\right) \right\}  \notag \\
&\triangleq &\min \{R_{1c}^{\max ,1}(P_{1p},P_{1c}),R_{1c}^{\max
,2}(P_{1p},P_{1c})\}  \notag \\
R_{1p}+2R_{1c} &\leq &\mathrm{C}\left( \frac{\beta
^{2}P_{1p}+2\alpha
^{2}P_{1c}}{1+2\alpha ^{2}P_{1p}}\right)  \label{cond2} \\
&\triangleq &R_{1}^{\mathrm{sum},1}(P_{1p},P_{1c})  \notag \\
R_{1p}+3R_{1c} &\leq &\mathrm{C}\left( \frac{\beta
^{2}P_{1p}+(2\alpha
^{2}+\beta ^{2})P_{1c}}{1+2\alpha ^{2}P_{1p}}\right)  \label{cond3} \\
&\triangleq &R_{1}^{\mathrm{sum},2}(P_{1p},P_{1c}).  \notag
\end{eqnarray}%
Notice that in writing the conditions (\ref{Rrs1_region}) we have removed
dominated inequalities.

In order to obtain some insight into the properties of the achievable rate
region of private and common messages $\mathcal{R}_{rs,1}(P_{1p},P_{1c})$
defined by inequalities (\ref{Rrs1_region}), Fig. \ref{achievableregions}
shows the region $\mathcal{R}_{rs,1}(P_{1p},P_{1c})$ for $P_{1p}=1,$ $%
P_{1c}=1,$ $\beta ^{2}=1$ and different values of $\alpha ^{2}.$ According
to the value of the inter-cell parameter $\alpha ^{2}$, the achievable
region $\mathcal{R}_{rs,1}(P_{1p},P_{1c})$ is a polyhedron with different
corner points. Fig. \ref{achievableregions} shows three illustrative cases
for small ($\alpha ^{2}=0.4$ in the figure), intermediate ($\alpha ^{2}=0.65$%
) and moderate inter-cell factor $\alpha ^{2}$ ($\alpha ^{2}=0.8$)\footnote{%
Notice that an exact determination of the threshold values of
$\alpha $ that lead to different regions is conceptually simple
but algebraically involved given the characterization
(\ref{Rrs1_region}). Moreover, we remark that we avoided the use
of the term strong ``interference" in this context in order
to be consistent with the conventional use of the term (see, e.g., \cite%
{kramer}).}$.$ In all cases, \textit{vertex A} has a simple interpretation
in terms of successive interference cancellation: in fact, it can be
achieved by first jointly decoding the common messages ($W_{c,m},$ $%
W_{c,m-1} $ and $W_{c,m+1})$, treating the private information as noise,
then cancelling the decoded common messages and finally decoding the
same-cell private message $W_{p,m}$. To show this, notice that, since in the
first decoding stage the channel seen by the three common messages at any RS
is a three-user MAC\ with noise power $1+(2\alpha ^{2}+\beta ^{2})P_{1p}$
(due to the interference from the primary messages), the common rate at
vertex A is given by $\min (R_{1c}^{1},R_{1c}^{2}),$ with
\end{subequations}
\begin{subequations}
\label{minR1R2}
\begin{eqnarray}
R_{1c}^{1}(P_{1p},P_{1c}) &=&\frac{1}{2}\mathrm{C}\left(
\frac{2\alpha
^{2}P_{1c}}{1+(2\alpha ^{2}+\beta ^{2})P_{1p}}\right)  \label{R1c1x} \\
R_{1c}^{2}(P_{1p},P_{1c}) &=&\frac{1}{3}\mathrm{C}\left(
\frac{(2\alpha ^{2}+\beta ^{2})P_{1c}}{1+(2\alpha ^{2}+\beta
^{2})P_{1p}}\right) . \label{R1c2x}
\end{eqnarray}

Our focus on vertex A in the achievable rate region $\mathcal{R}%
_{rs,1}(P_{1p},P_{1c})$ is justified by the following fact. Given the slope
of the side of the polyhedron $\mathcal{R}_{rs,1}(P_{1p},P_{1c})$ determined
by conditions (\ref{cond2})-(\ref{cond3}), it can be easily seen that
\textit{for each power allocation }$(P_{1p},P_{1c})$ \textit{vertex A
corresponds to the point where the rate on the first hop }$%
R_{rs,1}=R_{1p}+R_{1c}$ \textit{is maximum }and reads\textit{\ }
\end{subequations}
\begin{eqnarray}
R_{rs,1}^{\max }(P_{1p},P_{1c}) &=&R_{1p}^{\max }(P_{1p})+  \label{R1max} \\
&&+\min (R_{1c}^{1}(P_{1p},P_{1c}),R_{1c}^{2}(P_{1p},P_{1c})),  \notag
\end{eqnarray}%
with definitions (\ref{R1pmax}) and (\ref{minR1R2}). We remark the decoding
order that leads to vertex A (first common information, then private),
coupled with a specific power allocation, was recently shown in \cite{etkin}
to attain every point in the capacity region of the conventional
interference channel to within one bit. Finally, vertex points B and B$%
^{\prime }$ also have similar interpretations in terms of
successive interference cancellation. This is further discussed in
Appendix-A.

\textbf{Remark 1} (very strong interference): Similarly to the case of a
conventional interference channel \cite{carleial}, it can be shown that, if $%
\alpha ^{2}$ is sufficiently larger than the direct channel $\beta ^{2}$
(thus contradicting our assumption that $\alpha ^{2}\leq \beta ^{2}),$
transmission of only common messages ($P_{1p}=0$ and $P_{1c}=P_{1})$ is an
optimal strategy that is able to achieve the single-user upper bound to the
achievable rate, $R_{rs,1}=\log (1+\beta ^{2}P_{1}).$ The exact condition on
$\alpha ^{2}$ is derived in Appendix-B.

\subsection{Rate splitting in the second hop\label{sec_rs}}

With rate splitting in the first hop, each RS, say the $m$th, decodes in
each block the private message $W_{p,m}$ and the common message $W_{c,m}$ of
the same-cell MT$,$ along with the common messages of the adjacent cells $%
W_{c,m-1}$ and $W_{c,m+1}.$ The $m$th relay can then neglect the knowledge
of $W_{c,m-1}$ and $W_{c,m+1},$ and attempt to transmit to the $m$th BS the
two messages of the same-cell user $W_{p,m}$ and $W_{c,m}$ by using rate
splitting and interference cancellation exactly as explained in the previous
section for the first hop. Notice that the total rate $%
R_{rs,1}=R_{1p}+R_{1c},$ delivered to the RSs by the MTs, can be now split
into two streams, one private and one common, in a generally different share
with respect to the first hop. In particular, the signal transmitted by the $%
m$th RS is given by
\begin{equation}
Z_{m}=Z_{p,m}(V_{p,m})+Z_{c,m}(V_{c,m}),  \label{Zm_rs}
\end{equation}%
where $Z_{p,m}(\cdot )$ corresponds to a Gaussian codebook of rate $R_{2p}$
for the private message $V_{p,m}$ ($V_{p,m}\in \{1,2,...,2^{nR_{2p}}\}$) and
$Z_{c,m}(\cdot )$ is the $R_{2c}$-rate code for the common message $V_{c,m}$
($V_{c,m}\in \{1,2,...,2^{nR_{2c}}\}$)$.$ The total rate achievable on the
second hop thus becomes $R_{rs,2}=R_{2p}+R_{2c}.$ Moreover, the power
allocation is $P_{2}=P_{2p}+P_{2c},$ where $P_{2p}=E[|Z_{p,m}|^{2}]$ and $%
P_{2c}=E[|Z_{c,m}|^{2}]$. Similarly to the first hop, each BS is informed of
the private codebook used by the same-cell MT and of the common codebooks
employed by the same-cell MTs and the two adjacent-cell MTs.

Following the previous section, we can define the rate region $\mathcal{R}%
_{rs,2}(P_{2p},P_{2c})$ achievable in the second hop with rate
splitting for a given power allocation$.$ This is easily shown to
be defined by inequalities (\ref{Rrs1_region}), where subscript
``2" should \ be substituted for ``1" and parameters ($\gamma
^{2},\eta ^{2})$ should be written in lieu of ($\beta ^{2},\alpha
^{2})$. Accordingly, the maximum rate
in the second hop reads (recall (\ref{R1max}))%
\begin{eqnarray}
R_{rs,2}^{\max }(P_{2p},P_{2c}) &=&R_{2p}^{\max }(P_{2p})+  \label{Rrs2max}
\\
&&+\min (R_{2c}^{1}(P_{2c},P_{2c}),R_{2c}^{2}(P_{2p},P_{2c})),  \notag
\end{eqnarray}%
where $R_{2p}^{\max }(P_{2p}),$ $R_{2c}^{1}(P_{2c},P_{2c})$ and $%
R_{2c}^{2}(P_{2p},P_{2c})$ are obtained from (\ref{R1pmax}) and (\ref%
{minR1R2}), respectively, following the rules mentioned above.

Since with rate splitting in both hops the two hops are operated
independently, the optimal strategy is to transmit in both hops at the
maximum sum-rates $R_{rs,i}^{\max }(P_{ip},P_{ic})$ in (\ref{R1max}) and (%
\ref{Rrs2max}) for given power allocations $(P_{ip},P_{ic})$, $i=1,2$. It
follows that, optimizing over the power allocation on both hops, the rate
achievable with rate splitting in both hops is
\begin{equation}
R_{rs}=\underset{i=1,2}{\min }R_{rs,i}^{\max },  \label{Rrs}
\end{equation}%
with ($i=1,2$)
\begin{equation}
\begin{array}{c}
R_{rs,i}^{\max }=\underset{P_{ip},P_{ic}}{\max }R_{rs,i}^{\max
}(P_{ip},P_{ic}) \\
\text{s.t. }P_{ip}+P_{ic}=P_{i}.%
\end{array}
\label{rrs}
\end{equation}

\section{Improving the achievable rate in the second hop\label{sec_second}}

In this section, we investigate the performance of an alternative
transmission scheme for the second hop that leverages the common information
gathered at the RSs as a by-product of the use of rate splitting in the
first hop. This contrasts with the naive scheme discussed in Sec. \ref%
{sec_rs} whereby the common messages from adjacent cells were neglected when
transmitting in the second hop. Moreover, for reference, we evaluate the
rate achievable with rate splitting and multi-cell processing at the BSs (as
in the case where BSs are connected via a high capacity backbone) in Sec. %
\ref{sec_multicell}.

\subsection{Cooperative transmission at the relays\label{sec_coop}}

The rate splitting-based scheme discussed in Sec. \ref{sec_rs} for
transmission from RSs to BSs fails to exploit the knowledge of the common
messages of adjacent cells $W_{c,m-1}$ and $W_{c,m+1}$ at any $m$th RS.
Based on this side information, any $m$th cell could cooperate with the
adjacent cells $m-1$ (and $m+1$) in order to deliver the common message $%
W_{c,m-1}$ (and $W_{c,m+1})$ to the intended BS in cell $m-1$ (and $m+1$).
The presence of shared information among the transmitters has been
previously considered in the context of conventional ($2\times 2$)
interference channels in different scenarios. In particular, a model in
which the two transmitters have common information to deliver to \textit{both%
} receivers has been considered in \cite{maric1} \cite{jiang}, whereas an
asymmetric case where one transmitter has knowledge of the message of the
other transmitter was studied in \cite{maric2} \cite{wu} \cite{maric3}. Also
relevant is the case of a MAC\ channel with common information studied in
\cite{slepian}.

Similarly to the above mentioned works, here we adopt a superposition scheme
whereby transmitters cooperate for transmission of common information
towards the goal of achieving coherent power combining at the BSs. In
particular, the signal transmitted by the $m$th RS according to this scheme
is given by%
\begin{equation}
Z_{m}=Z_{p,m}(W_{p,m})+\sum_{i=-1}^{1}Z_{c,m+i}(W_{c,m+i}),  \label{Zm}
\end{equation}%
where $Z_{p,m}(\cdot )$ is defined as above and $Z_{c,m}(\cdot )$ accounts
for a common Gaussian codebook employed by the $m-1,$ $m$ and ($m+1)$th RSs
for cooperative relaying of the common messages $W_{c,m}.$ Notice that
variables $Z_{p,m}(\cdot )$ and $Z_{c,m}(\cdot )$ are uncorrelated. The
private ($W_{p,m}$) and common ($W_{c,m}$) messages are the ones sent in the
first hop by the MTs and therefore have rates $R_{1p}$ and $R_{1c},$
respectively. We focus on a simple power allocation among the transmitted
codewords in (\ref{Zm}), whereby the total power $P_{2}$ is divided as $%
P_{2}=P_{2p}+P_{2c}$ with $P_{2p}=E[|Z_{p,m}|^{2}]$ for the private part and
the power $P_{2c}$ equally shared among the three cooperative common signals
as $P_{2c}=3E[|Z_{c,m}|^{2}].$ Moreover, as in the previous section, each BS
is assumed to know the private codebook used by the same-cell MT and of the
common codebooks employed by the same-cell MTs and the two adjacent-cell
MTs. It should be remarked that a more general transmission scheme than the
one considered here could be employed where joint encoding of private\ $%
W_{p,m}$ and common $W_{c,m}$\ messages takes place at each $m$th RS
(instead of the independent encoding by which we interpret (\ref{Zm})),
similarly to \cite{slepian}. Here, for simplicity, we do not further pursue
the analysis of this scenario.

In order to derive the achievable rates of this scheme, let us substitute (%
\ref{Zm}) in the received signal (\ref{rx signal bs}) at the BSs (dropping
the arguments of the codewords):%
\begin{eqnarray}
Y_{m} &=&\gamma Z_{p,m}+(\gamma +2\eta )Z_{c,m}+(\gamma +\eta )Z_{c,m-1}+
\label{Ym} \\
&&+(\gamma +\eta )Z_{c,m+1}+S_{m}^{\prime }+M_{m},  \notag
\end{eqnarray}%
where $S_{m}^{\prime }$ represent the nuisance term due to the private
messages of adjacent cells and the common messages of cells $m-2$ and $m+2$:%
\begin{equation}
S_{m}^{\prime }=\eta Z_{p,m-1}+\eta Z_{p,m+1}+\eta Z_{c,m-2}+\eta Z_{c,m+2}.
\label{Sm}
\end{equation}%
We remark that the common messages of cells $m-2$ and $m+2$ ($Z_{c,m-2}$ and
$Z_{c,m+2}$) are considered as interference by the $m$th BS since they are
received without the benefit of cooperation from other RSs. Therefore,
adding the constraint of correct decoding of these messages at the $m$th BS
would reduce unnecessarily the rate $R_{1c}$ of the common codebooks $%
W_{c,i} $. From (\ref{Ym}), it can be seen that any $m$th BS observes a
four-user MAC\ channel with equivalent noise power $1+E[|S_{m}^{\prime
}|^{2}]=1+2\eta ^{2}(P_{2p}+P_{2c}/3).$ Therefore, similarly to Sec. \ref%
{sec_rs_region}, the achievable rates ($R_{1p},P_{1c}$) of the private and
common information belong to the rate region $\mathcal{R}%
_{coop,2}(P_{2p},P_{2c})$ characterized by:
\begin{eqnarray*}
R_{1p} &\leq &\mathrm{C}\left( \frac{\gamma ^{2}P_{2p}}{1+2\eta
^{2}(P_{2p}+P_{2c}/3)}\right) \\
R_{1c} &\leq &\min \left\{ \frac{1}{2}\mathrm{C}\left(
\frac{2(\gamma +\eta
)^{2}P_{2c}}{1+2\eta ^{2}(P_{2p}+P_{2c}/3)}\right) ,\right. \\
&&\left. \frac{1}{3}\mathrm{C}\left( \frac{(2(\gamma +\eta
)^{2}+(\gamma
+2\eta )^{2})P_{2c}}{1+2\eta ^{2}(P_{2p}+P_{2c}/3)}\right) \right\} \\
R_{1p}+2R_{1c} &\leq &\mathrm{C}\left( \frac{\gamma
^{2}P_{2p}+2(\gamma
+\eta )^{2}P_{2c}}{1+2\eta ^{2}(P_{2p}+P_{2c}/3)}\right) \\
R_{1p}+3R_{1c} &\leq &\mathrm{C}\left( \frac{\gamma
^{2}P_{2p}+(2(\gamma
+\eta )^{2}+(\gamma +2\eta )^{2})P_{1c}}{1+2\eta ^{2}(P_{2p}+P_{2c}/3)}%
\right) .
\end{eqnarray*}

The maximum achievable rate with rate splitting in the first hop and
cooperative transmission in the second hop, according to the coding scheme
described above, can be found by solving the following optimization problem:
\begin{eqnarray}
R_{coop} &=&\max_{R_{1p},R_{1c},P_{1p},P_{1c},P_{2p},P_{2c}}R_{1p}+R_{1c}
\label{Rcoop} \\
&&\text{s.t. }\left\{
\begin{array}{l}
P_{ip}+P_{ic}=P_{i},\text{ }i=1,2 \\
(R_{1p},R_{1c})\in
\begin{array}{c}
\mathcal{R}_{rs,1}(P_{1p,}P_{1c})\cap \\
\mathcal{R}_{coop,2}(P_{2p,}P_{2c}).%
\end{array}%
\end{array}%
\right.  \notag
\end{eqnarray}%
Notice that for each choice of the power allocation $%
(P_{1p,}P_{1c},P_{2p,}P_{2c})$, the optimization problem (\ref{Rcoop}) can
be solved by linear programming.

\subsection{Multi-cell processing\label{sec_multicell}}

In this section we consider the possibility of performing joint decoding of
the received signals at the BSs \cite{somekh-review}. As mentioned above,
this requires the presence of a high capacity backbone connecting all the
BSs to a central processor. We assume the use of rate splitting in the first
hop, whereas in the second hop the cooperative transmission scheme of Sec. %
\ref{sec_coop}, which aims at coherent power combining at the BSs for the
common messages, is employed.

Similarly to \cite{wyner}, we can interpret the received signal (\ref{Ym})-(%
\ref{Sm}) as an equivalent inter-symbol interference (ISI) channel over the
BSs:%
\begin{equation}
Y_{m}=h_{p,m}\ast Z_{p,m}+h_{c,m}\ast Z_{c,m}+M_{m},  \label{convolution}
\end{equation}%
where ``$\ast$" denotes convolution and the finite-impulse response filters $%
h_{nc,m}$ and $h_{c,m}$ are given by
\begin{subequations}
\label{channels}
\begin{eqnarray}
h_{p,m} &=&\eta \delta _{m+1}+\gamma \delta _{m}+\eta \delta _{m-1} \\
h_{c,m} &=&\eta \delta _{m+2}+(\gamma +\eta )\delta _{m+1}+(\gamma +2\eta
)\delta _{m} \\
&&+(\gamma +\eta )\delta _{m-1}+\eta \delta _{m-2},  \notag
\end{eqnarray}%
with $\delta _{m}$ denoting the Kronecker delta function ($\delta _{m}=1$
for $m=0$ and $\delta _{m}=0$ elsewhere). The channel (\ref{convolution})-(%
\ref{channels}) is a Gaussian MAC with ISI \cite{cheng-verdu} so that,
allocating the transmission powers as in Sec. \ref{sec_coop}, the region $%
\mathcal{R}_{mcp,2}(P_{2p,}P_{2c})$ of achievable rates ($R_{1p},R_{1c}$) in
the second hop with multicell processing and relay cooperation is easily
shown to satisfy the following conditions:
\end{subequations}
\begin{eqnarray*}
R_{1p} &\leq &\int_{0}^{1}\mathrm{C}\left( P_{2p}(\gamma +2\eta
\cos (2\pi
f))^{2}\right) df \\
R_{1c} &\leq &\int_{0}^{1}\mathrm{C}\left( \frac{P_{2c}}{3}(\gamma
+2\eta
+2(\gamma +\eta )\cos (2\pi f)+\right. \\
&&\left. +2\eta \cos (4\pi f))^{2}\right) df \\
R_{1p}+R_{1c} &\leq &\int_{0}^{1}\mathrm{C}\left( P_{2p}(\gamma
+2\eta \cos
(2\pi f))^{2}\right. \\
&&+\frac{P_{2c}}{3}(\gamma +2\eta ++2(\gamma +\eta )\cos (2\pi f)+ \\
&&\left. +2\eta \cos (4\pi f))^{2}\right) df.
\end{eqnarray*}%
Finally, accounting for both first and second hops, the rate achievable with
rate splitting, relay cooperation and multicell processing can be obtained
by solving the following optimization problem:
\begin{eqnarray}
R_{mcp} &=&\max_{R_{1p},R_{1c},P_{1p},P_{1c},P_{2p},P_{2c}}R_{1p}+R_{1c}
\label{Rmcp} \\
&&\text{s.t. }\left\{
\begin{array}{l}
P_{ip}+P_{ic}=P_{i},\text{ }i=1,2 \\
(R_{1p},R_{1c})\in
\begin{array}{c}
\mathcal{R}_{rs,1}(P_{1p,}P_{1c})\cap \\
\mathcal{R}_{mcp,2}(P_{2p,}P_{2c}).%
\end{array}%
\end{array}%
\right.  \notag
\end{eqnarray}%
Notice again that, for fixed power allocation $%
(P_{1p,}P_{1c},P_{2p,}P_{2c}), $ problem (\ref{Rmcp})\ can be solved by
linear programming. As a final remark, we recall that, as stated in Sec. \ref%
{sec_coop}, an alternative transmission scheme to (\ref{Zm}) could employ
joint encoding of common and private messages following \cite{slepian}. The
performance advantages of this solution are not further investigated here.

\section{Numerical results\label{sec_numerical}}

\begin{figure}
\centering
 \includegraphics[scale=0.5]{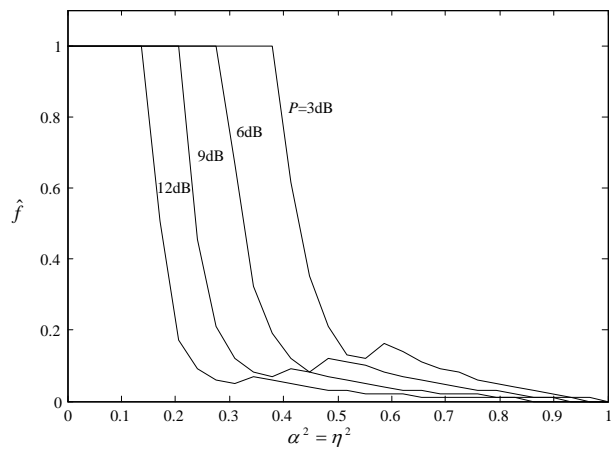}
 \caption{Optimal fraction $\hat{f}$ of power devoted to the transmission of
private messages when rate splitting is used in both hops versus
inter-cell
gains $\protect\alpha ^{2}=\protect\eta ^{2}$ ($\protect\beta ^{2}=\protect%
\gamma ^{2}=1).$}
 \label{optf}
\end{figure}

Here we present some numerical results\ in order to corroborate the analysis
and gain some insight into the performance of the proposed coding schemes.
Throughout this section, we set $\beta ^{2}=\gamma ^{2}=1$ and $\alpha
^{2}=\eta ^{2}.$ We are interested at first in investigating the conditions
under which rate splitting is advantageous over single-rate transmission.
Toward this goal, we consider a symmetric scenario with $P_{1}=P_{2}=P$ and
evaluate the optimal fraction of power $f$ to be devoted to the private
message assuming rate splitting in both hops as per (\ref{rrs}). By
symmetry, it is clear that the optimal fraction $\hat{f}$ is the same in
both hops, i.e., $\hat{f}=\hat{P}_{1p}/P=\hat{P}_{2p}/P,$ where the hat
notation identifies optimal quantities. Fig. \ref{optf} shows the optimal
fraction $\hat{f}$ versus the inter-cell gains $\alpha ^{2}=\eta ^{2}.$ It
can be seen that for small inter-cell gains $\alpha ^{2}=\eta ^{2},$ it is
optimal to use single-rate transmission ($\hat{f}=1)$ until a given
threshold gain, after which it is in general increasingly better to devote
more power to common messages. This result is in line with the known results
on the interference channel \cite{carleial} \cite{han} and confirms our
initial motivation (see Sec. \ref{sec_transmission}). Moreover, for
increasing power $P$ the threshold gain at which common messages should
carry more power decreases significantly.

\begin{figure}
\centering
 \includegraphics[scale=0.5]{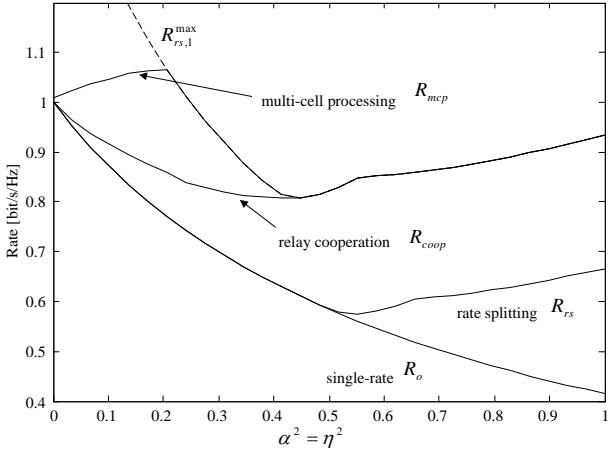}
 \caption{Achievable rates with single-rate transmission
 $R_{o}$ (\protect\ref{Ro}), rate splitting in both hops $R_{rs}$
(\protect\ref{Rrs}), relay cooperation in the second hop
$R_{coop}$ (\protect\ref{Rcoop}) and multi-cell processing in the
second hop $R_{mcp}$ (\protect\ref{Rmcp}) versus inter-cell gains
$\protect\alpha ^{2}=\protect\eta ^{2}$ ($P_{2}=0.5\cdot P_{1},$
$P_{1}=3dB).$}
 \label{rate3dB}
\end{figure}

\begin{figure}
\centering
 \includegraphics[scale=0.51]{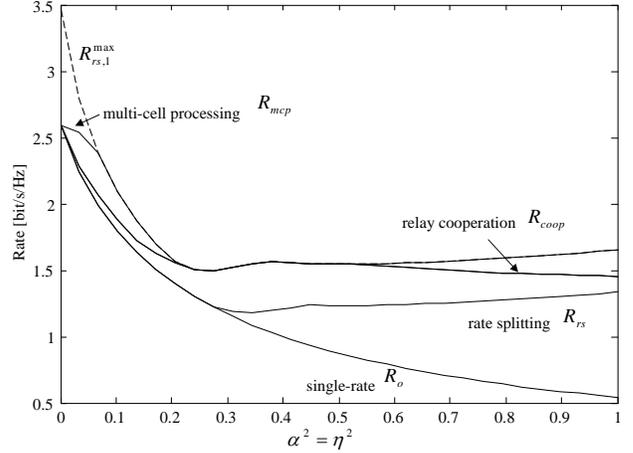}
 \caption{Achievable rates with
single-rate transmission $R_{o}$ (\protect\ref{Ro}), rate
splitting in both hops $R_{rs}$ (\protect\ref{Rrs}), relay
cooperation in the second hop $ R_{coop}$ (\protect\ref{Rcoop})
and multi-cell processing in the second hop $ R_{mcp}$
(\protect\ref{Rmcp}) versus inter-cell gains
$\protect\alpha^{2}=\protect\eta ^{2}$ ($P_{2}=0.5\cdot P_{1},$
$P_{1}=10dB).$}
 \label{rate6dB}
\end{figure}

We now turn to the performance assessment of rate splitting (with possible
cooperation or multi-cell processing in the second hop) in terms of
achievable rates. In order to obtain meaningful results, we focus on a
scenario where the second hop is the bottleneck by setting $P_{2}=P_{1}/2$
(to be interpreted in linear scale)$.$\ While this might not be the case in
typical applications where RSs are fixed and endowed with a power supply, it
is an interesting case study to assess the possible benefits of more
elaborate processing in the second hop. Figs. \ref{rate3dB} and \ref{rate6dB}
show the achievable rates with single-rate transmission $R_{o}$ (\ref{Ro}),
rate splitting in both hops $R_{rs}$ (\ref{Rrs}), cooperation at the relays
in the second hop $R_{coop}$ (\ref{Rcoop}) and multi-cell processing in the
second hop $R_{mcp}$\ (\ref{Rmcp}) for $P_{1}=3dB$ and $P_{1}=10dB$,
respectively. Also shown is the maximum rate achievable on the first hop
with rate splitting and optimal power allocation $R_{rs,1}^{\max }$ (\ref%
{rrs}). This provides an upper bound on the overall achievable rate in the
considered scenario where the second hop creates the performance bottleneck.
It can be seen that: (\textit{i}) as expected from the discussion on Fig. %
\ref{optf}, rate splitting is advantageous with respect to single-rate
transmission if the inter-cell gains $\alpha ^{2}=\eta ^{2}$ are large
enough; (\textit{ii}) for sufficiently small signal-to-noise ratio (i.e.,
power $P_{1})$ cooperation at the relays provides relevant performance gains
over rate splitting in both hops and allows to achieve the upper bound $%
R_{rs,1}^{\max }$ for $\alpha ^{2}=\eta ^{2}$ large enough (Fig. \ref%
{rate3dB}); (\textit{iii}) for signal-to-noise-ratio sufficiently large, the
additional interference created by the common messages relayed with
cooperative transmission in the second hop (recall the discussion in Sec. %
\ref{sec_coop}) has a deleterious effect on the rate if gains $\alpha
^{2}=\eta ^{2}$ are relevant and, accordingly, the benefits of cooperation
are less pronounced (Fig. \ref{rate6dB}); (\textit{iv}) multi-cell
processing in the second hop allows the system to achieve the upper bound $%
R_{rs,1}^{\max }$ for $\alpha ^{2}=\eta ^{2}$ large enough.

\section{Conclusions}

In a mesh network with a regular (cellular) structure, there exists a rich
structure in the underlying wireless connections that can be exploited in
order to design more effective coding strategies. In this paper, we have
explored one such opportunity for a two-hop mesh network with one active
user (and relay) per cell. In particular, we have exploited the presence of
meaningful inter-cell propagation paths (from terminals to relays and/or
from relays to base stations) by considering the use of a rate splitting
coding approach, which is know to be close to optimal (or even optimal, in
certain cases) for conventional interference channels. Based on this basic
scheme, we have further proposed an alternative cooperative transmission
scheme in the second hop, that takes advantage of the side information
available at the relays as a by-product of the use of rate splitting in the
first hop. Numerical results confirm that rate splitting is able to provide
significant gains as long as the inter-cell power gains are large enough.

\section{Appendix}

\subsection{Further discussion on the capacity regions in Fig. \protect\ref%
{achievableregions}}

In Sec. \ref{sec_rs_region}, the successive interference strategy achieving
the rate-maximizing vertex A in the rate region $\mathcal{R}%
_{rs,1}(P_{1p},P_{1c})$ was discussed in detail (recall Fig. \ref%
{achievableregions}). Here we would like to further interpret the corner
points B and B$^{\prime }$ in terms of successive interference cancellation.
Vertex B, arising in scenarios with weak interference, is obtained by
detecting first the common message from same-cell MT, then the private
message from same-cell MT and finally common messages from adjacent-cell.
This leads to $R_{1c}=R_{1c}^{\max ,1}$ and $R_{1p}=R_{1p}^{1}=\mathrm{C}%
\left( \frac{\beta ^{2}P_{1p}}{1+2\alpha ^{2}P_{1p}+2\alpha ^{2}P_{1c}}%
\right) .$ Similarly, vertex B$^{\prime },$ arising with intermediate
interference, can be achieved by first detecting the private message and
then jointly recovering the common messages, leading to $R_{1c}=R_{1c}^{\max
,2}$ and $R_{1p}=R_{1p}^{2}=\mathrm{C}\left( \frac{\beta ^{2}P_{1p}}{%
1+2\alpha ^{2}P_{1p}+(2\alpha ^{2}+\beta ^{2})P_{1c}}\right) $.
Finally, vertex C is characterized by the common rate
$R_{1c}^{3}=\mathrm{C}\left( \frac{\beta ^{2}P_{1c}}{1+2\alpha
^{2}P_{1p}+\beta ^{2}P_{1p}+2\alpha ^{2}P_{1c}}\right) .$

\subsection{Derivation of the condition of very strong interference}

Following Remark 1, here we look for conditions on the inter-cell power gain
$\alpha ^{2}$ that allow that system to achieve the single-user upper bound $%
R_{rs,1}=\mathrm{C}(\beta ^{2}P_{1})$ on the achievable rate of
the first hop, through transmission of only common messages.
Setting $P_{1c}=P_{1}$ (and $P_{1p}=0$)$,$ we need to impose the
condition that all the rate\ inequalities defining the capacity
region of the three-user MAC channel seen
by the common messages at each RS support rates larger than $\mathrm{C}%
(\beta ^{2}P_{1}).$ Notice that, since here we allow $\alpha ^{2}>\beta
^{2}, $ we should now consider all the seven inequalities of the MAC\
capacity region (as opposed to (\ref{Rrs1_region}) where some bounds were
dominated under the assumption that $\alpha ^{2}\leq \beta ^{2}).$ This
leads to: (\textit{i}) from single-user bounds, it immediately follows that
we need $\alpha ^{2}\geq \beta ^{2};$ (\textit{ii}) from two-user bounds, we
have
\begin{subequations}
\begin{eqnarray}
\frac{1}{2}\mathrm{C}(2\alpha ^{2}P_{1}) &\geq &\mathrm{C}(\beta
^{2}P_{1})
\\
\frac{1}{2}\mathrm{C}((\alpha ^{2}+\beta ^{2})P_{1}) &\geq &\mathrm{C}%
(\beta ^{2}P_{1}),
\end{eqnarray}%
from which we obtain
\end{subequations}
\begin{equation}
\alpha ^{2}\geq \beta ^{2}\cdot \max \left( \frac{P_{1}}{2}+1,\beta
^{2}P_{1}+1\right) ;  \label{a1}
\end{equation}%
(\textit{iii}) from three-user bounds, it follows that
\begin{equation}
\frac{1}{3}\mathrm{C}((2\alpha ^{2}+\beta ^{2})P_{1})\geq
\mathrm{C}(\beta ^{2}P_{1}),
\end{equation}%
which implies%
\begin{equation}
\alpha ^{2}\geq \beta ^{2}\cdot (2+3P_{1}+\beta ^{4}P_{1}).  \label{a2}
\end{equation}%
Noticing that condition (\ref{a2}) dominates (\ref{a1}) for any $\beta ^{2},$
we finally obtain the result that, in order for rate-splitting to achieve
the single-user bound, we need an inter-cell power gain that satisfies the
very strong interference conditions (\ref{a2}).

\end{document}